\documentclass[journal, a4paper]{IEEEtran}

\usepackage{dblfloatfix} 
\usepackage{lipsum,booktabs}
\usepackage{comment}
\usepackage{amsmath,amssymb,amsfonts}
\usepackage{algorithmic}
\usepackage{mathtools, cuted}
\usepackage{graphicx}
\usepackage{hyperref}

\usepackage{orcidlink}
\usepackage{textcomp}
\usepackage{xcolor}
\usepackage{balance}
\usepackage{enumitem}

\usepackage{pifont}
\def\passed{\color{green!40!black}\ding{52}}
\def\failed{\color{red}\ding{55}}

\newcommand{\softname}[1]{\textsc{#1}}
\usepackage{enumitem}

\usepackage{doi}

\usepackage{colortbl}
\newcommand{\cellcolorgray}[1]{\cellcolor{gray!25}{#1}}

\usepackage{multirow}
\usepackage{booktabs}
\usepackage[numbers]{natbib}
\usepackage{url}
\usepackage{lipsum}
\usepackage{makecell}
\usepackage{algorithm}
 \usepackage{graphicx} 
\usepackage{longtable}
\usepackage[lofdepth,lotdepth]{subfig}

\AtBeginDocument{%
\providecommand\BibTeX{{%
    \normalfont B\kern-0.5em{\scshape i\kern-0.25em b}\kern-0.8em\TeX}}}

\usepackage{multicol}
\usepackage{fancyvrb}

\newcommand{\nameMM}{RSMM}
\newcommand{\focusa}[1]{\textbf{#1}}
\newcommand{\practice}[1]{\emph{#1}}
\newcommand{\capability}[1]{\textbf{#1}}

\newcommand{\submissionnote}[0]{Submitted to IEEE eScience 2024 Conference}

\title{\nameMM{}: A Framework to Assess Maturity of Research Software Project
\thanks{\submissionnote{}}}

\usepackage{authblk}

\author[1,2]{Deekshitha\orcidlink{0000-0003-1831-8941}}
\author[2]{Rena Bakhshi\orcidlink{0000-0002-2932-3028}}
\author[2]{Jason Maassen \orcidlink{0000-0002-8172-4865}}
\author[2]{Carlos Martinez Ortiz\orcidlink{0000-0001-5565-7577}}
\author[3]{\\Rob van Nieuwpoort\orcidlink{0000-0002-2947-9444}}
\author[1]{Slinger Jansen \orcidlink{0000-0003-3752-2868}}

\affil[1]{Utrecht University, Utrecht, The Netherlands \and
   \{d.deekshitha,slinger.jansen\}@uu.nl}
\affil[2]{Netherlands eScience Center, Amsterdam, The Netherlands
 \{r.bakhshi, j.maassen, c.martinez\}@esciencecenter.nl}
\affil[3]{Leiden University, Leiden, The Netherlands r.v.van.nieuwpoort@liacs.leidenuniv.nl}

\begin{document}

  
\maketitle

\begin{abstract}
The organizations and researchers producing research software face a common problem of making their software sustainable beyond funding provided by a single research project. This is addressed by research software engineers through building communities around their software, providing appropriate licensing, creating reliable and reproducible research software, making it sustainable and impactful, promoting, and ensuring that the research software is easy to adopt in research workflows, etc. As a result, numerous practices and guidelines exist to enhance research software quality, reusability, and sustainability. However, there is a lack of a \emph{unified framework} to systematically integrate these practices and help organizations and research software developers refine their development and management processes. Our paper aims at bridging this gap by introducing a novel framework: \nameMM{}. It is designed through systematic literature review and insights from interviews with research software project experts. In short, \nameMM{} offers a structured pathway for evaluating and refining research software project management by categorizing 79 best practices into 17 capabilities across 4 focus areas. From assessing code quality and security to measuring impact, sustainability, and reproducibility, the model provides a complete evaluation of a research software project maturity. With \nameMM{}, individuals as well as organizations involved in research software development gain a systematic approach to tackling various research software engineering challenges. By utilizing \nameMM{} as a comprehensive checklist, organizations can systematically evaluate and refine their project management practices and organizational structure.

\end{abstract}
\vspace{0.35cm}




\section{Introduction}
Open science is a movement that strives for greater openness and collaboration in research practices. It promotes the open sharing of publications, data, software, and various academic outputs as early as possible, making them accessible for reuse. As a result, open science leads to greater scientific and societal impact~\cite{OS2024}. Research software plays an important role in the open science movement~\cite{heumuller2020publish}. It includes source code files, algorithms, scripts, computational workflows, and executables created during the research process or for a research purpose~\cite{fair4wg}. For good scientific practice, the resulting research software should be open and adhere to the FAIR principles~\cite{barker2022introducing,Fortunato-open-21} to allow repeatability, reproducibility, and reuse. Compared to research data, research software should be both archived for reproducibility and actively maintained for reusability~\cite{hasselbring2020fair,fairseco}. 

Traditionally, researchers focus on maximizing their production process towards scientific publications rather than the quality and sustainability of their research software. This is due to the pressure on researchers to continually publish their work to advance their careers (also known as \emph{publish-or-perish})~\cite{pp17}. 
Long term sustainability strategies such as software quality and community building require resources in terms of time and expertise. Researchers may lack the necessary skills and do not prioritize these activities because these efforts are often not acknowledged by research organisations and funding agencies~\cite{incenstives-2023}. For the same reasons, the peer review process, which serves as a quality check for scientific publications and in open source software development, is rarely done for research software.

Over the past few years, however, the significance of research software has grown exponentially with its increased usage~\cite{barker_michelle_2021_5762703,carver-2022}. 
The Research Software Engineering (RSE) movement, which started in the UK, has experienced substantial growth and rapid emergence~\cite{8994167}. This movement aims to recognize the vital importance of software for research, and the role of people, policy, and infrastructure in its development, support, and maintenance~\cite{lamprecht2022we}.
As a result, currently there is significant interest from researchers, RSEs and even funding organizations in establishing best practices for research software engineering~\cite{martinez_ortiz_carlos_2020_4310217, barker2022introducing} and to ensure research software becomes sustainable, reproducible, and has a community around it. However, while numerous practices and guidelines exist,  
there is a lack of a \emph{unified framework} to systematically integrate these practices and assist researchers and organizations in refining their development and management processes.

We present the Research Software focus area Maturity Model (\nameMM{}), targeted at organizations that produce research software. Our model aims to provide a structured analysis of research software engineering practices that can help researchers and research supporting organizations produce research software. 
\nameMM{} takes the FAIR principles as the baseline but goes further into software sustainability, community building, the software role in reproducibility, usability, and other quality factors relevant to the open science movement. This model includes 4 focus areas: \focusa{Software Project Management}, \focusa{Research Software Management}, \focusa{Community Engagement}, and \focusa{Software Adoptability}. Each focus area covers several best practices to improve its capabilities. \nameMM{} facilitates an incremental pathway for organizations to enhance their research software project management. By utilizing \nameMM{} as a comprehensive checklist, organizations can systematically evaluate and refine their project management practices and organizational structure.

The contributions of this study are as follows: 
\begin{itemize}
\item \textbf{Maturity model for research software:} We present \nameMM{} to assess the maturity of research software projects by considering best practices of research software project management. The framework includes 79 best practices covering code quality and security, sustainability, reproducibility, community building, and many others required to improve the research software project management. 
\item \textbf{Community-based approach:} The framework is designed based on interviews with researchers, research software engineers, and project managers. 
\item \textbf{The published dataset:} 
The model itself, along with a complete description of the 79 practices, is publicly available as a dataset~\cite{RSMM-dataset}. This description includes when the practices are implemented and is organized based on the MoSCoW prioritization (Must have, Should have, Could have, Won't have). Additionally, it details the resources required for execution, dependencies among neighboring practices, and references.
\end{itemize}
The remainder of the paper is structured as follows:
Section \ref{background} presents maturity models and a comparative analysis between our model and the existing ones.
In Section \ref{research_method}, we provide an in-depth exploration of the research methodology employed in designing \nameMM{}. Section \ref{model} explains our novel model \nameMM{} and Section \ref{discussion} illustrates the method of assessing research software project maturity using  \nameMM{}, supported by two examples.
Section~\ref{conculsion} concludes the paper by summarizing the findings and contributions made in this work. 

\section{Background and related work}\label{background}

\subsection{Maturity Models Concept}
Maturity is an evolutionary progress in demonstrating a specific ability or
accomplishing a target from an initial stage to a desired end stage. 
Maturity models capture this process from an initial state to the desired end state and can guide organizations in assessing and developing organizational capabilities~\cite{poeppelbuss2011maturity}. 
Maturity models are tools developed for organizations to evaluate and compare, providing a basis for improvement and informed strategies to enhance specific areas within the organization~\cite{de2005understanding}. 

To reach a particular maturity level, an organization must meet specific criteria and characteristics related to capabilities or process performance. The initial stage represents a starting point where the organization may have limited capabilities in the investigated domain. Organizations with the most advanced capabilities reach the highest maturity level~\cite{jansen2020focus, overeem2022api,van2010design}.
This maturity level assessment of capabilities helps derive and prioritize improvement measures.
The following subsection compares \nameMM{} with the existing models.


\subsection{Landscape of Maturity models}

Few existing maturity models are used for software capability management but are not specific to the research software development. Therefore, they do not cover aspects such as sustainability, reproducibility, impact measurement, promotion, visibility and adoptability.

Capability Maturity Model Integration (CMMI) and its predecessor Capability Maturity Model (CMM) are industry standard maturity models~\cite{CMMI3}. They include the 5 maturity levels. CMM is used to assess an organization software development processes in terms of maturity. It helps developers to enhance software quality and the overall software development process.
CMMI v3.0~\cite{CMMI3} goes even beyond software development and includes process quality assurance, configuration management, monitor and control, planning, estimating
requirements development and management governance,
implementation infrastructure,
organizational training, process management
verification and validation and etc. Therefore, it considers developers and other departments such as marketing, finance, and purchasing. This broader scope might be deemed unnecessarily complex when applied to Agile software development practices~\cite{patel2009agile} like Extreme Programming, SCRUM, and Lean development are typical for research software projects.

\begin{figure*}
\includegraphics[page=1, clip,trim=0cm 0.2cm 0cm 0cm, width=1\textwidth]{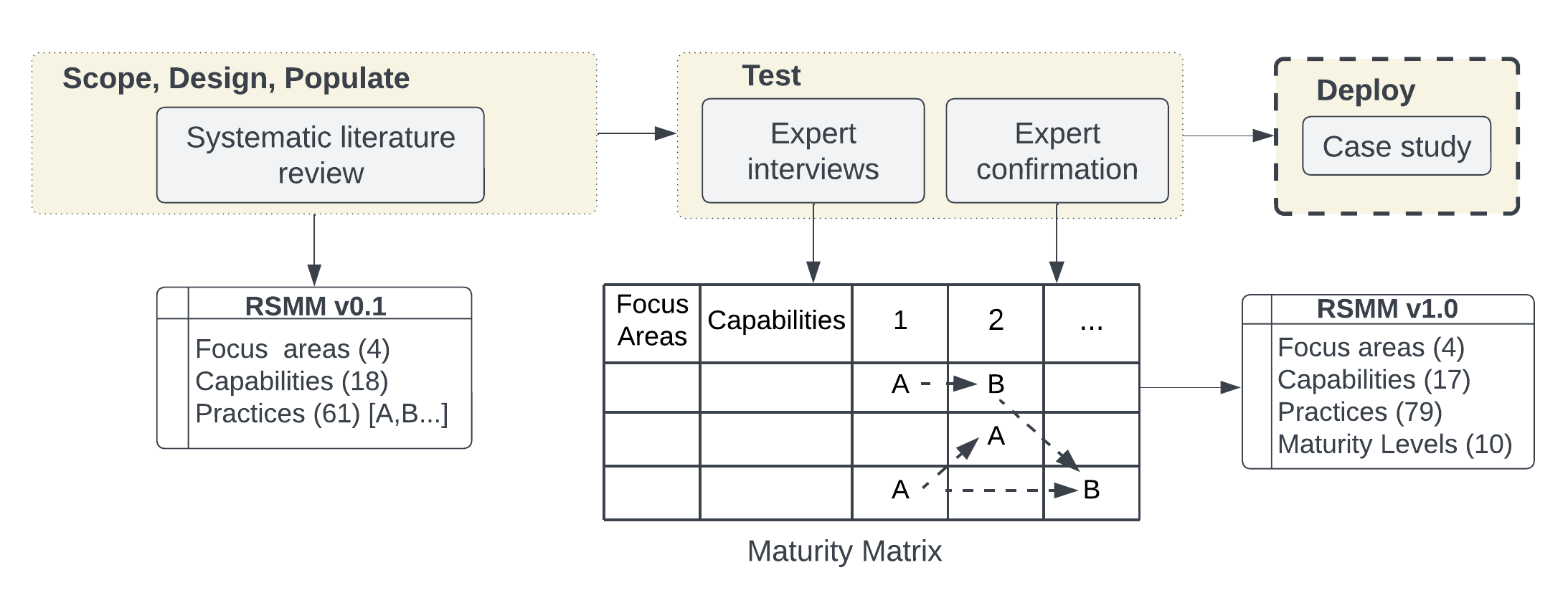}
\vspace*{-0.5cm}
 \caption{Design phases of \nameMM{}: The steps involved in developing a \nameMM{} v1.0: In the \emph{Scope}, \emph{Design}, and \emph{Populate} phases, we conducted a systematic literature review to collect and explore academic and grey literature, resulting in the creation of \nameMM{} v0.1. In the \emph{Test} phase, we included Expert  Interviews and Expert Confirmations. Each stage in this phase leads to the evolution of \nameMM{}, also producing intermediate versions v0.2 and v0.3 (available upon request from authors). Case studies are conducted in the \emph{Deploy} phase to validate the model applicability.}
\label{fig:focusareaMM}
\end{figure*}

The focus area maturity model (FAMM) is one type of maturity model~\citep{van2010design}. It helps organizations to measure their performance in a particular functional domain. A functional domain consists of different focus areas, each with its capabilities. These capabilities are arranged in a maturity matrix, which helps to identify different maturity levels. Each capability includes various improvement actions, known as practices, that support the organization in gradually improving in that functional domain.
Unlike other maturity models, FAMM does not have fixed maturity levels; maturity levels can start from 0 and end at any positive integer. Each focus area is evaluated separately and has its maturity levels. The most relevant examples of FAMM are:

SEG-M\textsuperscript{2}~\cite{jansen2020focus} is a focus area maturity model designed to enhance the governance of software ecosystems within organizations. The model consists of 168 practices collected from structured literature reviews and desk studies, following the maturity model 
creation steps outlined by de Bruin et al.~\cite{de2005understanding}. 
However, SEG-M$^2$ focuses on many aspects of software ecosystem governance, such as ecosystem health, open markets, and market and sale-related capabilities. Thus, these are mainly applicable to industry-based software projects. 

API-m-FAMM~\cite{overeem2022api} is a model designed for managing APIs. The study integrates De Bruin et al. methodology with Design Science Research~\cite{de2005understanding}, using the card sorting technique and multiple evaluation cycles. This model includes capabilities such as version management, documentation, community engagement, and resource management. 

Many maturity models exist for the management process of open-source projects~\cite{opensource,paulk1993capability,crawford2006project}. While there are many similarities between those models and ours, a significant difference is that researchers commonly use different criteria for assessing open-source research software (e.g., reproducibility, visibility, impact). Furthermore, open-source projects focus more on testing, reputation, and generic integration. In contrast, \nameMM{} focuses on various aspects such as sustainability, visibility, impact measurement~\cite{fairseco}, integration of research software into research workflows, adoptability, fostering partnerships, improving developers' skill sets, and recognizing their contributions to research software projects. This model is designed from the perspectives of users, developers, funders, and partners, giving equal importance to all these stakeholders. As we already mentioned, our dataset is publicly available~\cite{RSMM-dataset} and includes a complete description of all the model components. This dataset can assist individuals even with a limited software engineering training to understand the practices and resources required to implement them.   

In summary, most of these models or frameworks serve industry-level projects; however, none specifically focus on research software project management. This managemnet process includes community building, visibility, conference promotion, and more. Another area for improvement is that many of these models are limited to a particular domain or capability. \nameMM{} is designed to address these existing problems by focusing on the research software project management and the number of capabilities within a single framework.

To the best of our knowledge, \nameMM{} is the first research software maturity model that combines the best practices from software development and open-source software development while giving importance to the research software engineering practices. To this end, the focus of \nameMM{} extends beyond the developmental aspects of research software, giving equal importance to non-code considerations such as building a community around the research software and enhancing its impact, sustainability, and adoption within the research community.

\section{Designing the Maturity Model}\label{research_method}
Inspired by the previous FAMMs~\cite{jansen2020focus, overeem2022api}, we have developed our model \nameMM{}. This section covers its design phases.

\begin{figure*}
\centering
 \includegraphics[page=1, clip, width=1.0\linewidth,trim=0cm 0cm 0cm 0cm,]
{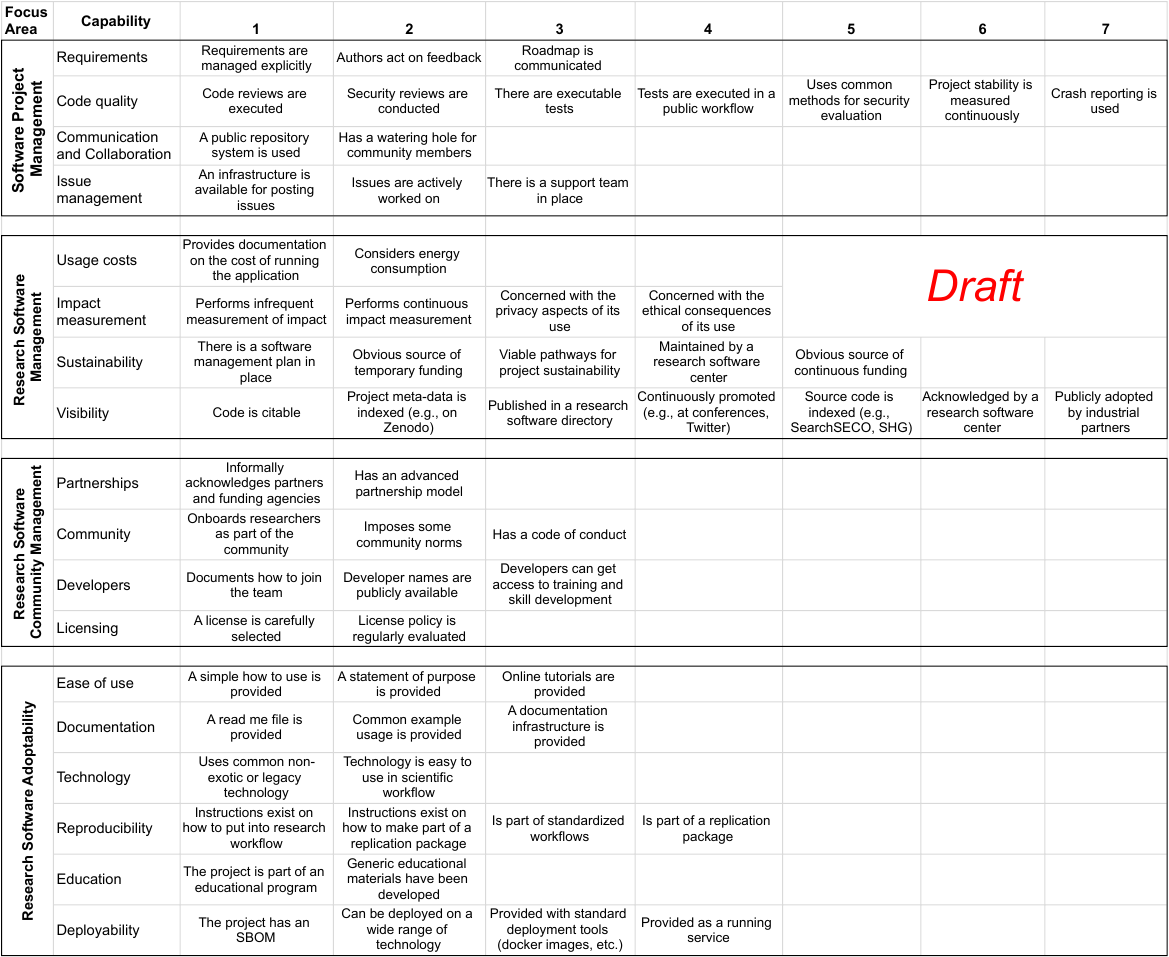} %
 \vspace{-0.5cm}
 \caption{\nameMM{} v0.1 ({\color{gray}not the final model}): The initial model includes 4 focus areas, 18 capabilities, and 61 practices. We included the practices collected through the systematic literature review from both academic and grey literature in the model, with maturity levels from 1 to 7. These practices are not placed based on their maturity.}
\label{fig:fammmetamodel}
\end{figure*}

\subsection{Research Method}\label{famm_model}
We used De Bruin's 5-phase approach~\cite{de2005understanding} to design \nameMM{} with the following phases (see Fig.~\ref{fig:focusareaMM}):
\begin{itemize}
    \item \textbf{Scope:} The scope of \nameMM{} is to evaluate and improve research software project management.
\item \textbf{Design:} The design phase focuses on the questions "why," "how," and "who", as outlined below: 
\begin{itemize}
    \item The \textbf{Why}: The purpose of \nameMM{} is to help an organization that produces research software to improve their research software project management by assessing and improving the maturity of their projects. 
    \item The \textbf{How}: \nameMM{} acts as a guideline and helps organizations to understand and implement the best practices effectively to achieve their desired maturity level.
    \item The \textbf{Who}: The intended audience of \nameMM{} are researchers, research software engineers, research software project managers, funders and policy makers.
\end{itemize}
\item \textbf{Populate:} We identified the focus areas, capabilities, and practices of research software project management through a systematic literature review, resulting in \nameMM{} v0.1. Section \ref{literaturereview} explains the literature review process to collect components of \nameMM{}.
\item \textbf{Test:}
We conducted expert interviews during this phase to position practices within the maturity matrix. Furthermore, we sent the resulting model, \nameMM{} v1.0, to the interview experts for confirmation.
Section \ref{expert_interview} describes how the collected practices are positioned based on their maturity in the matrix (maturity matrix).
\item \textbf{Deploy:} 
In this phase, as part of the future work, we will 
 conduct comprehensive case study to validate and verify the applicability of the model. Section \ref{discussion} discusses how to use this model to evaluate research software project management on two examples.
 \end{itemize}
 
The following subsections briefly explain the \emph{Populate}, \emph{Test} and \emph{Deploy} phases.

\subsection{Literature Review: Scope, Design, and Populate}\label{literaturereview}
As discussed in Section~\ref{background}, existing methods for evaluating and improving research software project management need to be revised.
Initially, we conducted a systematic literature review gathering academic and grey literature. This process helps us to identify the practices, capabilities, and focus areas of research software project management to design the model. We utilized  snowballing method (also known as backward and forward snowballing), that is used to expand a literature review by identifying articles relevant to the topic of interest~\cite{wohlin2014guidelines}. 
We conducted our literature review using the ACM Digital Library and Google Scholar. We used the search string, \textit{``(``software management'' OR ``software project management'') AND (``research software'' OR ``research software project management'') AND (``community development'' OR ``community engagement'' OR ``community participation'') AND (``research software'' OR ``scientific software'') AND (``FAIR principles'' OR ``best practices'')''} to collect articles related to our topic. Our search was limited to the past 10 years and yielded 36 results in total from both databases. After removing duplicates and irrelevant entries, we retained 19 academic papers and grey literature. 
Next, we look at older papers' references to find more relevant sources, adding 5 more papers. After that, we found newer documents that cite the original ones to get the latest insights. We then identified practices, capabilities, and focus areas from these collected documents. Based on these findings, we started a second round of searching, this time focusing specifically on the identified capabilities and focus areas. As a result of this refined search, 66 new sources (both academic papers and grey literature) were added to the pool of resources. Then, we grouped practices into capabilities and capabilities into focus areas and vice versa (following both a top-down approach and bottom-up approach for grouping and collecting focus areas, capabilities, and practices~\cite{de2005understanding}).

From this method, we identified 61 best practices and grouped these practices into capabilities and focus areas. The described process is a part of the \emph{Populate} phase, which results in 4 focus areas, 18 capabilities, and 61 practices (indicated as A, B in Figure \ref{fig:focusareaMM}). As depicted in the figure, these practices and capabilities together form the focus areas that represent the functional domain of the research software project management. 
In the model, \nameMM{}, the capabilities define the ability to achieve a goal related to the research software project management by executing two or more interrelated practices. 
In this case, we can define a practice as an action need to be taken to improve the research software project management. As it is shown in Fig.~\ref{fig:fammmetamodel}, we placed practices of \nameMM{} v0.1 without considering its maturity. Next, we followed 12 expert interviews to determine the positioning of these practices. The following we provide details about this expert interview.

\subsection{Expert Interviews and confirmation: Test Phase}\label{expert_interview}

The \emph{Test} phase of the design methodology includes semi-structured expert interviews and confirmation steps for designing \nameMM{} v1.0. We consulted experts in research software development and project management to find the positioning of the practices in the maturity matrix and validate the completeness and correctness of the model. We followed 5 steps in the \emph{test} phase, described in details below:

\subsubsection{Connecting with experts through Dutch and German RSE groups}
We initiated the interview process by approaching experts through Dutch and German RSE groups, providing a brief project overview, and inviting them to express their interest in participating. Participation in the interview 
process was entirely voluntary.

\subsubsection{Expert selection}
We received 15 positive responses. Following this, we shared the interview protocol with these interested participants, resulting in 12 confirmations from individuals who met the interview selection criteria of our study. We set specific criteria for expert selection to ensure participants had the experience and knowledge in research software project management to obtain valuable results. The expert selection criteria are as follows:
\begin{enumerate}[label=\arabic*)]
\item Participants must indicate they are knowledgeable
on a minimum of two out of the 4 focus areas of the
\nameMM{};
\item Participants must have a minimum of 3 years
of experience in either developing or managing research software projects;
\item Participants must work at an organization as a Research Software Engineer, researcher, project manager as a part of a team working on research software projects or any comparable role related to research software.
\end{enumerate}
The shared interview protocol outlines procedural steps, rules, and regulations, including the questions to address during the interview.
Additionally, the file includes a description of the focus areas,
capabilities, and practices used in the model.

\subsubsection{Conducting expert interviews} We scheduled online meetings based on the
participants' availability. The interviews took between 45 minutes and 1 hour and 36 minutes. 
Before the interview, we shared our maturity model without including maturity levels. During the expert interview, we asked the experts to arrange practices into the corresponding capability groups based on the order of maturity, seeking their insights.

\begin{figure}[ht]
\centering
\includegraphics[width=1.0\linewidth,trim=0cm 0cm 0cm 0cm,]{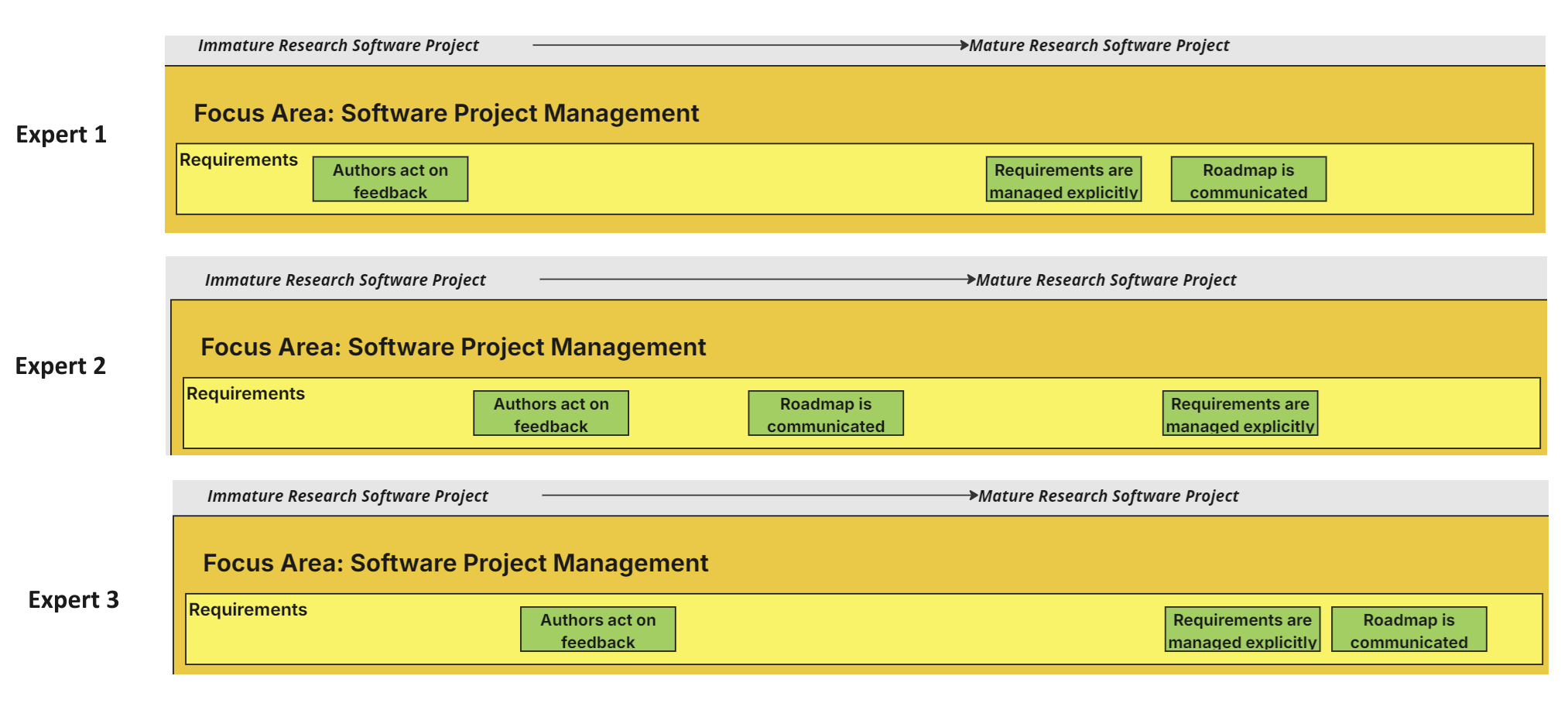}
\vspace{-2em}
\caption{Maturity Model template by experts: The screenshot depicts 3 experts' ranking of practices within the \capability{Requirements} capability, illustrating their placements in the \nameMM{} v0.1a expert's dashboard.}
\label{fig:expert_template}
\end{figure}

After completing this task for each focus area, we ask for suggestions about any missing capabilities or practices within that specific focus area and whether any practices were misclassified into the wrong capability. This systematic approach is repeated for all 4 focus areas. Figure~\ref{fig:expert_template} depicts a comparison of the maturity model templates provided by the 3 experts. The ranking of the first capability \capability{Requirements} is presented according to the inputs provided by the 3 experts. The final placement of practices is determined by considering the dependencies between different practices  
(these are shown in Fig.~\ref{fig:fammmetamodel} as arrow marks). For example, the practice, \practice{Provide executable tests} must be implemented or executed before the \practice{Execute tests in a public workflow}. We refer interested reader to the dataset~\cite{RSMM-dataset} for the detailed information.

Upon completion of this task,  we asked experts questions to evaluate the model applicability, ease of use, and feasibility.

\subsubsection{Interview data processing}
We mapped practices in each expert's dashboard to maturity levels, then identified the positioning of these practices by taking the mode of all experts response. For example, out of 12 experts, 4 recommended positioning the \practice{Provide executable tests} practice at level 4. Consequently, we assigned this practice to maturity level 4, considering its dependencies with another practice, \practice{Execute tests in a public workflow}. We revised the practice names to follow the 'Verb Object (Qualifier)' pattern. We also made small adjustments to the names of focus areas and capabilities.

\subsubsection{Expert confirmation}
We incorporated the suggestions from the interviewees into a new version of the model, \nameMM{} v1.0. During the interviews, experts added 18 new practices. We then validated the new model by sending it back to all interviewees for a second round of feedback. We are still awaiting final feedback from all the experts, but one of the first comments we received was \textit{``Yes, I will use input from this model in my future work''}, which we perceive as encouraging.

In total, we received the approval from 8 experts to publish \nameMM{} v1.0, including a few suggestions to modify the dataset. These suggestions will be considered in future versions of the model.

\subsection{Evaluation of research software projects: Deploy Phase}\label{deploy}
We assessed the applicability of \nameMM{} v1.0 by evaluating a couple of research software projects in the \emph{Deploy} phase. Section \ref{discussion} presents the evaluation results for research software projects \softname{GGIR}~\cite{van_hees_vincent_2022_7043054} and \softname{ESMValTool}~\cite{Andela_ESMValTool_2023}. 

\section{Our Maturity Model}\label{model}
\begin{figure}[!t]
\centering
\includegraphics[width=1.0\linewidth,trim=0cm 0cm 0cm 0cm]{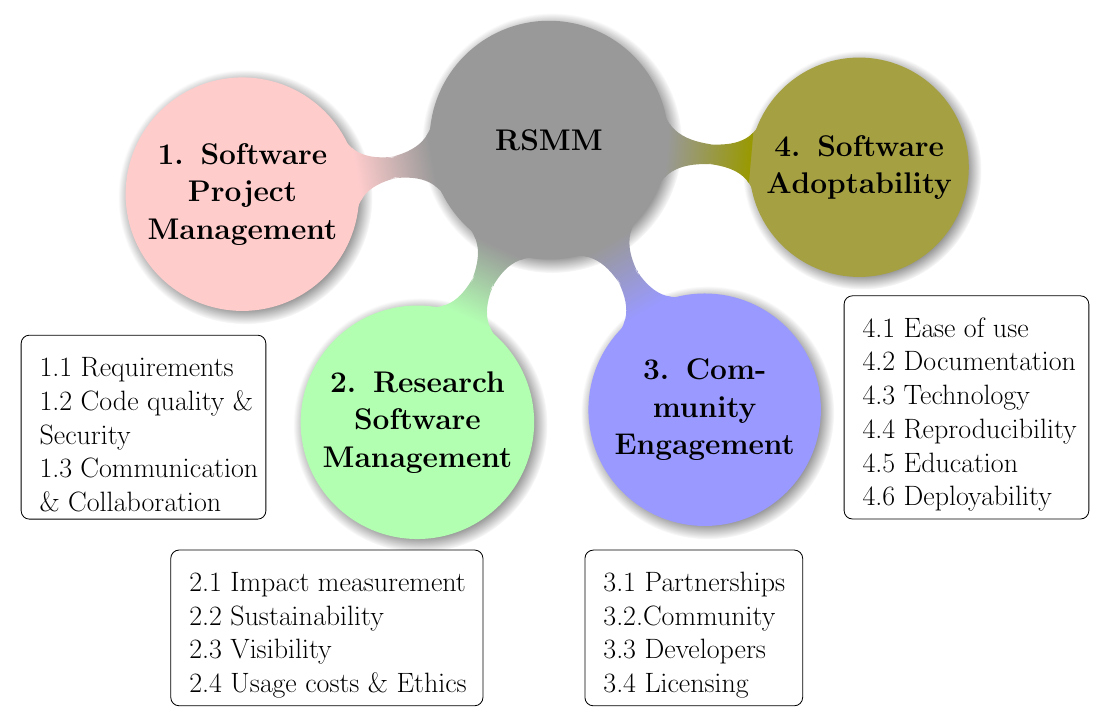}
\vspace{-1.5em}
\caption{\nameMM{} Focus areas and capabilities: The 4 focus areas and 17 capabilities of the \nameMM{} v1.0 are shown in the figure.}
\label{fig:famm_focus}
\end{figure}

\begin{figure*}
\hspace{-1cm}
\includegraphics[page=1, clip,trim=0cm 10cm 0cm 2cm,width=1.1\textwidth]{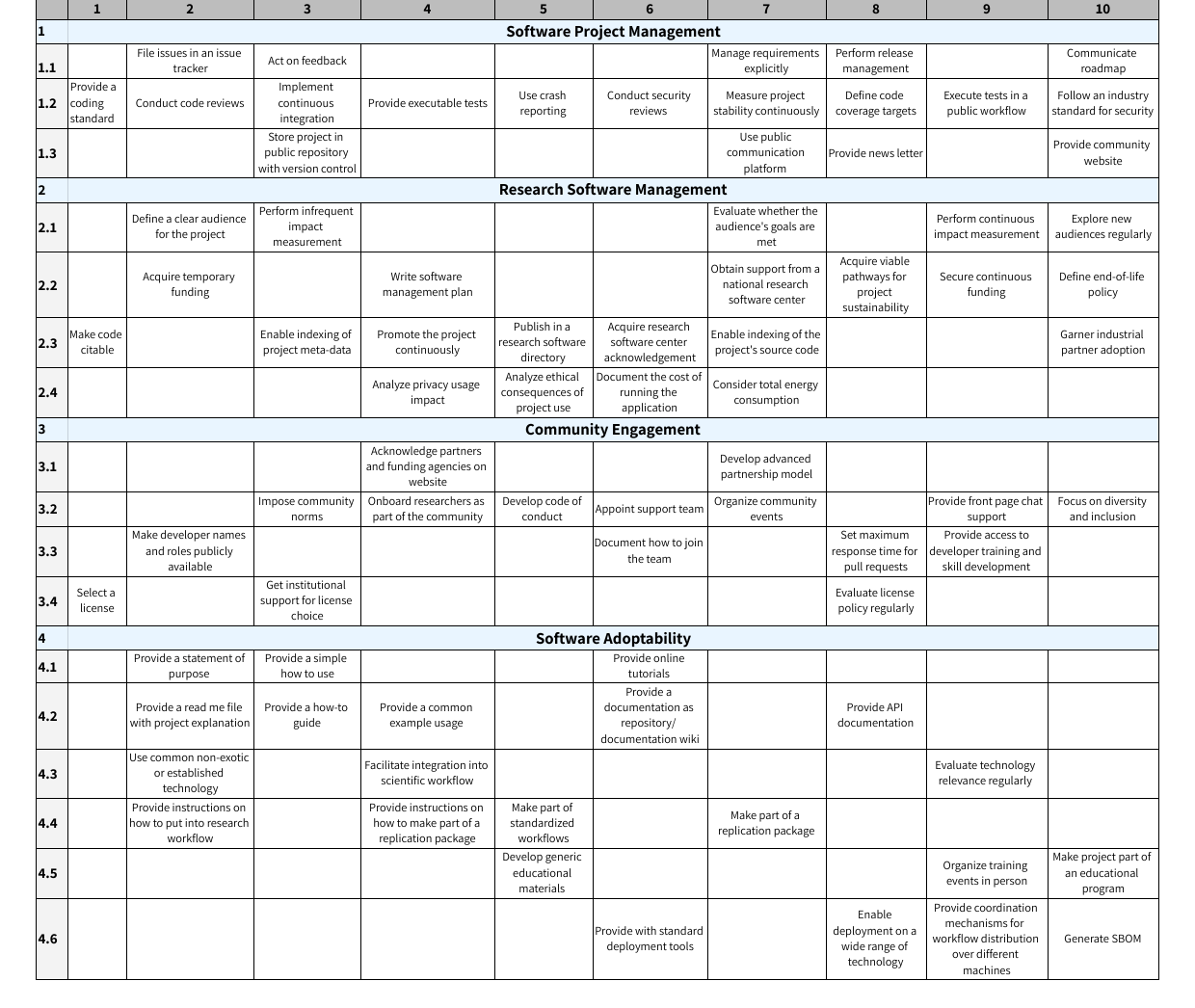}
 \caption{\nameMM{} v1.0: The updated and final version of \nameMM{}. It includes 4 focus areas, 17 capabilities, and 79 practices. These practices are placed between maturity levels 1 to 10. The first column of the table includes the capability codes.}
\label{fig:fammmetamodelv1.0}
\end{figure*}

This section presents the resulting model \nameMM{} v1.0, which includes 4 focus areas related to Research Software Project Management: \focusa{Software Project Management}, \focusa{Research Software Management}, \focusa{Community Engagement}, and \focusa{Software Adoptability}. Figure \ref{fig:famm_focus} illustrates the focus areas, their associated capabilities, and their respective code names. The following subsection briefly describes 4 focus areas of the \nameMM{} v1.0, and this model is depicted in Fig.~\ref{fig:fammmetamodelv1.0}.

\subsection{Software project management}
\focusa{Software project management} manages the resources and work activities needed to develop and modify software-intensive systems~\cite{10.5555/1074100.1074810}. The software project is a highly
people-intensive effort that extends over a considerable period, significantly impacting the work and performance of various stakeholders, including project managers. The
primary success criteria for software managers are delivery of systems that satisfy specified needs and \capability{requirements}, on time and within budget~\cite{10.5555/1074100.1074810}. 
Software development involves challenges, including evolving technology, immature technology, sloppy development practices, and staff changes. \focusa{Software project management} addresses these complexities by promoting stakeholder involvement, managing risks, and fostering transparent communication, helping project managers navigate and overcome these obstacles.
 
Thus, various factors influence the software project management process, and they are almost the same for research software development. We included 3 capabilities and 19 practices related to this focus area. The dataset consists of descriptions of all the practices.

 

\subsection{Research software management}
Like software management, \focusa{Research software management} is a process of managing research software. Increasing usage of research software in the various research domains highlights the importance of adhering to best practices~\cite{barker2022introducing}. In this focus area, we included 4 capabilities such as \capability{Impact measurement}, \capability{Sustainability}, \capability{Visibility}, and \capability{Usage cost and Ethics} and 22 practices.

The 3 practices of \nameMM{} v1.0 for improving research software visibility are \practice{Make code citable}, \practice{Enable indexing of project meta-data}~\cite{barker2022introducing}, and \practice{Publish in a research software directory}~\cite{Spaaks_Research_Software_Directory_2020}. Figure \ref{fig:example_practice} describes the practice \practice{Publish in a research software directory}. The practice code (2.3.5) is generated by combining the focus area (2), capability (3), and maturity level (5). The description for when this practice is implemented is categorized into MoSCoW categories (with W omitted), including the resources required to complete it, its dependencies on other practices, and references provided within the practice description set.

\begin{figure}
\noindent
\fbox{%
    \parbox{\dimexpr\columnwidth-1\fboxsep-1\fboxrule\relax}%
{
\noindent\textbf{Practice Code:} 2.3.5 \newline
\noindent\textbf{Practice Name:}  Publish in a research software directory
\newline
\noindent\textbf{Description:} Publishing research software projects in a research software directory facilitates discovery, citation, and collaboration among researchers, research software engineers, and users within the research community.
\newline
\noindent\textbf{When implemented:} 
\begin{itemize}
\item
(M) Identifying and selecting suitable research software directories that align with the project focus area, target audience, and visibility goals, considering factors such as reputation, coverage, and accessibility.
\item(M) Preparation and submission of comprehensive meta-data and documentation for the research software project to the selected directory.
\item(S) Compliance with directory-specific submission guidelines, metadata standards, and Quality Assurance criteria to ensure accurate representation, indexing, and visibility of the research software project within the directory database and search interface.
\item(C) Monitoring and updating of project listings and meta-data in the research software directory as needed to reflect changes, updates, or new releases of the software.
\end{itemize}
\noindent\textbf{Resources required:} 
\begin{itemize}
\item
Time: Required for researching and selecting appropriate research software directories, preparing and submitting project listings, and maintaining directory entries over time. Ongoing time and effort may be needed to monitor directory performance, respond to inquiries, and update project information.
\item
Knowledge or expertise in directory submission processes, metadata standards, and documentation requirements for research software projects
\item
Collaboration with research software directory team to publish projects on their website.
\end{itemize}
\noindent\textbf{Dependencies:}  \newline
\noindent\textbf{References:} ~\cite{ Spaaks_Research_Software_Directory_2020}
}
} 
\caption{Description set for the practice \practice{Publish in a research software directory}.}
\vspace{-0.5cm}
\label{fig:example_practice}
\end{figure}

\subsection{Community Engagement}
A community generally evolves and maintains research software, creating an ecosystem of competing and collaborative products. It is influenced by the open-source movement culture of sharing and collaboration~\cite{8565942}. Consequently, we have identified 4 capabilities and 16 practices associated with this focus area to foster community development around research software. Below, we describe one of these practices.

\practice{Develop code of conduct}: A code of conduct outlines how participants should communicate, enforces consequences for violations, and reflects the community values. It creates an inclusive space where everyone can contribute comfortably, regardless of gender, ethnicity, or sexual orientation~\cite{tourani2017code}. Thus, a code of conduct helps effective collaboration, leading to successful project deliverables.

\subsection{Software Adoptability}
The focus area \focusa{Software Adoptability} 
concerns with how easily and effectively 
research software can be adopted and utilized by users. This focus area is aimed at  understanding and enhancing the user-friendliness, accessibility, and overall adoption strategies of research software by the research community.
It includes 6 capabilities and 22 practices. Below are descriptions of one capability and one of its practices.

\capability{Documentation}: Good documentation is needed for the reuse of research software ~\cite{hermann2022documenting}. For example, including examples in the documentation (\practice{Provide a common example usage}) offers users a starting point for experimentation. When examples demonstrate the software functionality, users can quickly execute the code and enhance their understanding. All practices within this \capability{Documentation} capability, viewed from the user perspective, emphasize the importance of clarity and accessibility in helping users effectively understand and reuse any research software project.

\begin{table*}[!htb]
\scriptsize
\centering
\subfloat[][The maturity level of \softname{GGIR} is 4-3-6-7]{
\begin{tabular}{|l|l|l|l|l|l|l|l|l|l|l|l|}
 \hline
\textbf{\begin{tabular}[c]{@{}l@{}}Focus \\ Area\end{tabular}} & \textbf{\begin{tabular}[c]{@{}l@{}}Capa- \\ bility\end{tabular}}  & \textbf{1} & \textbf{2} & \textbf{3} & \textbf{4} & \textbf{5} & \textbf{6} & \textbf{7} & \textbf{8} & \textbf{9} & \textbf{10}
\\ \hline
\multirow{3}{*}{\textbf{1}} & \textbf{1.1}  & \cellcolorgray{} & \cellcolorgray{} & \cellcolorgray{} \passed& \cellcolorgray{} &&&
\passed & \passed & & \failed
   \\ \cline{2-12}
 & \textbf{1.2} 
& \cellcolorgray{} \passed & \cellcolorgray{}\passed & \cellcolorgray{}\passed &
\cellcolorgray{}\passed & \failed & \failed &
\passed & \failed &
\failed & \failed 
   \\ \cline{2-12}
 & \textbf{1.3}  & \cellcolorgray{}& \cellcolorgray{}& \cellcolorgray{}\passed &\cellcolorgray{}
&&&\passed & \failed & &
\failed 
   \\ \hline
\multirow{3}{*}{\textbf{2}} & \textbf{2.1}  & \cellcolorgray{}&
\cellcolorgray{}\passed & \cellcolorgray{}\passed & &&&
\failed & & \passed & \failed 
   \\ \cline{2-12}
 & \textbf{2.2}  &\cellcolorgray{}& \cellcolorgray{}\passed &\cellcolorgray{}& \failed & && \passed &
\failed & \failed &\failed
   \\ \cline{2-12}
 & \textbf{2.3} & \cellcolorgray{}\passed & \cellcolorgray{}& \cellcolorgray{}\passed & \passed & \passed &
\passed & \passed & & & \passed 
\\ \cline{2-12}
 & \textbf{2.4}  & \cellcolorgray{} & \cellcolorgray{}& \cellcolorgray{} & \passed & \failed & \failed & \failed & &
   &\\  \hline
\multirow{4}{*}{\textbf{3}} & \textbf{3.1}  &\cellcolorgray{}&\cellcolorgray{}&\cellcolorgray{}& \cellcolorgray{}\passed &\cellcolorgray{} &\cellcolorgray{}& \failed & &&
\\ \cline{2-12}
 & \textbf{3.2}  &\cellcolorgray{}&\cellcolorgray{}& \cellcolorgray{}\passed &
\cellcolorgray{}\passed & \cellcolorgray{}\passed & \cellcolorgray{}\passed &
\failed & & \failed & \failed
   \\ \cline{2-12}
 &  \textbf{3.3}  &\cellcolorgray{}& \cellcolorgray{}\passed &
\cellcolorgray{}&\cellcolorgray{}&\cellcolorgray{}& \cellcolorgray{}\passed & & \failed & \failed &
   \\ \cline{2-12}
 & \textbf{3.4} & \cellcolorgray{}\passed &
\cellcolorgray{}& \cellcolorgray{}\passed & \cellcolorgray{}&\cellcolorgray{}&\cellcolorgray{}&& \failed & &
   \\ \hline
\multirow{6}{*}{\textbf{4}} &
  \textbf{4.1} &\cellcolorgray{}& \cellcolorgray{}\passed & \cellcolorgray{}\passed &\cellcolorgray{}&\cellcolorgray{}&\cellcolorgray{}
  \passed &\cellcolorgray{}&&&
   \\ \cline{2-12}
 & \textbf{4.2}  &\cellcolorgray{} & \cellcolorgray{}\passed & \cellcolorgray{}\passed & \cellcolorgray{}\passed &\cellcolorgray{}
& \cellcolorgray{}\passed &\cellcolorgray{} & \failed & & \\ \cline{2-12}
& \textbf{4.3}  & \cellcolorgray{}& \cellcolorgray{}\passed &\cellcolorgray{} &
\cellcolorgray{}\passed &\cellcolorgray{} &\cellcolorgray{} &\cellcolorgray{} & & \failed & \\ \cline{2-12}
& \textbf{4.4} &\cellcolorgray{} &\cellcolorgray{} \passed &\cellcolorgray{} & \cellcolorgray{}\passed &\cellcolorgray{} \passed & \cellcolorgray{}& \cellcolorgray{}\passed & & &  \\ \cline{2-12}
& \textbf{4.5}  &\cellcolorgray{}&\cellcolorgray{}&\cellcolorgray{}&\cellcolorgray{}&\cellcolorgray{}
\passed &\cellcolorgray{}&\cellcolorgray{}&&\passed &\passed
   \\ \cline{2-12}
 & \textbf{4.6}  & \cellcolorgray{} &\cellcolorgray{} &\cellcolorgray{} &\cellcolorgray{}&\cellcolorgray{}&\cellcolorgray{}
   \passed &\cellcolorgray{} & \passed & \failed & \failed \\ \hline
\end{tabular}
}
\subfloat[][The maturity level of \softname{ESMValTool} is 5-4-8-8.]{
\begin{tabular}{|l|l|l|l|l|l|l|l|l|l|l|l|}
 \hline
\textbf{\begin{tabular}[c]{@{}l@{}}Focus \\ Area\end{tabular}} & \textbf{\begin{tabular}[c]{@{}l@{}}Capa- \\ bility\end{tabular}}  & \textbf{1} & \textbf{2} & \textbf{3} & \textbf{4} & \textbf{5} & \textbf{6} & \textbf{7} & \textbf{8} & \textbf{9} & \textbf{10}
\\ \hline
\multirow{3}{*}{\textbf{1}} & \textbf{1.1}  & \cellcolorgray{} & \cellcolorgray{}\passed  & \cellcolorgray{}\passed &\cellcolorgray{}  &\cellcolorgray{} & & \passed
& \passed & & \passed 
   \\ \cline{2-12}
 & \textbf{1.2} 
& \cellcolorgray{}\passed & \cellcolorgray{}\passed &\cellcolorgray{} \passed & \cellcolorgray{}\passed
 & \cellcolorgray{}\passed  & \failed & \passed
 & \passed & \passed
 &  \failed 
   \\ \cline{2-12}
 & \textbf{1.3}  &\cellcolorgray{} & \cellcolorgray{} & \cellcolorgray{}\passed & \cellcolorgray{}
&\cellcolorgray{} & &\passed & \passed & & \passed

   \\ \hline
\multirow{3}{*}{\textbf{2}} & \textbf{2.1}  & \cellcolorgray{}&
\cellcolorgray{} & \cellcolorgray{}\passed  & \cellcolorgray{} \passed &&& \passed
& &   \passed & \failed
   \\ \cline{2-12}
 & \textbf{2.2}  &\cellcolorgray{}& \cellcolorgray{}\passed &\cellcolorgray{}& \cellcolorgray{}\passed & && \passed &
\passed & \passed & \failed
   \\ \cline{2-12}
 & \textbf{2.3} &\cellcolorgray{} \passed & \cellcolorgray{}& \cellcolorgray{}\passed & \cellcolorgray{}\passed& \passed &
\passed & \passed & & & \failed
\\ \cline{2-12}
 & \textbf{2.4}  & \cellcolorgray{}  & \cellcolorgray{}& \cellcolorgray{}& \cellcolorgray{}\passed & \failed & \failed & \failed& &
   &\\  \hline
\multirow{4}{*}{\textbf{3}} & \textbf{3.1}  & \cellcolorgray{}& \cellcolorgray{}& \cellcolorgray{}&\cellcolorgray{}\passed &\cellcolorgray{}&\cellcolorgray{}& \cellcolorgray{}\passed&\cellcolorgray{} &&
\\ \cline{2-12}
 & \textbf{3.2}  &\cellcolorgray{}&\cellcolorgray{}& \cellcolorgray{}\passed  &\cellcolorgray{}\passed
 &\cellcolorgray{}\passed  & \cellcolorgray{}\passed &\cellcolorgray{}\passed
 &\cellcolorgray{} & \failed & \failed
   \\ \cline{2-12}
 &  \textbf{3.3}  &\cellcolorgray{} &\cellcolorgray{}\passed  &\cellcolorgray{}
&\cellcolorgray{}&\cellcolorgray{}&\cellcolorgray{}\passed &\cellcolorgray{} & \cellcolorgray{}\passed & \passed &
   \\ \cline{2-12}
 & \textbf{3.4} & \cellcolorgray{}\passed &\cellcolorgray{}
&\cellcolorgray{}\passed &\cellcolorgray{} &\cellcolorgray{}&\cellcolorgray{}&\cellcolorgray{}& \cellcolorgray{}\passed  & &
   \\ \hline
\multirow{6}{*}{\textbf{4}} &
  \textbf{4.1} &\cellcolorgray{}& \cellcolorgray{}\passed & \cellcolorgray{}\passed& \cellcolorgray{}& \cellcolorgray{}& \cellcolorgray{}\passed &\cellcolorgray{}&\cellcolorgray{}&&
   \\ \cline{2-12}
 & \textbf{4.2}  & \cellcolorgray{}& \cellcolorgray{}\passed & \cellcolorgray{}\passed &\cellcolorgray{}\passed &\cellcolorgray{}
&\cellcolorgray{}\passed  & \cellcolorgray{}&\cellcolorgray{} \passed & &  \\ \cline{2-12}
& \textbf{4.3}  & \cellcolorgray{}& \cellcolorgray{}\passed  &\cellcolorgray{}&
\cellcolorgray{}\passed &\cellcolorgray{} &\cellcolorgray{}&\cellcolorgray{}&\cellcolorgray{} & \failed & \\ \cline{2-12}
& \textbf{4.4} &\cellcolorgray{}&\cellcolorgray{}\passed&\cellcolorgray{}& \cellcolorgray{}\passed  & \cellcolorgray{}\passed&\cellcolorgray{} &\cellcolorgray{}\passed  &\cellcolorgray{} & &  \\ \cline{2-12}
& \textbf{4.5}  &\cellcolorgray{}&\cellcolorgray{}&\cellcolorgray{}&\cellcolorgray{}& \cellcolorgray{}\passed
&\cellcolorgray{}&\cellcolorgray{}&\cellcolorgray{}& \passed & \failed
   \\ \cline{2-12}
 & \textbf{4.6}  &\cellcolorgray{}  & \cellcolorgray{}& \cellcolorgray{}& \cellcolorgray{}& \cellcolorgray{}& \cellcolorgray{}\passed
  &\cellcolorgray{} & \cellcolorgray{}\passed&  \failed&\failed \\ \hline
  
\end{tabular}
}
\caption{Case study: the evaluation results of the research software \softname{GGIR} (left) and \softname{ESMValTool} (right) using \nameMM{} v1.0. The tick mark in the practices box indicates that the tool follows those practices, while the cross mark indicates that it did not. Grey shading is added to the cells to show the longest path of each focus area that achieves maturity.}
\label{tab:evaltable}
\end{table*} 

\section{Validation and Discussion}\label{discussion}
In this section, we illustrate on 2 case studies how our model can be used to evaluate research software project management. 
%
Such an evaluation can help researchers and RSEs identifying the next steps for improving the maturity of their own research software. It assists them in choosing existing software for adoption, and allows them to track the progress and improvements over time. For research organisations and funding agencies, our model can be used to assess the maturity and viability of research software projects, thereby allowing them to make an informed decision on which projects to support and fund.



For our case study, we selected 2 research software projects: \softname{GGIR} and \softname{ESMValTool}.
\softname{GGIR} is an R-package to process and analysis multi-day data collected with wearable raw data accelerometers for physical activity and sleep research. \softname{ESMValTool} is a community-developed climate model diagnostics and evaluation software package.
We collected data related to \softname{GGIR} and \softname{ESMValTool} from their websites and GitHub repositories. Based on this data, we evaluated both research software projects using \nameMM{} v1.0, resulting in maturity scores of 4-3-6-7 for \softname{GGIR} and 5-4-8-8 for \softname{ESMValTool} across the corresponding focus areas. The details of these evaluations are presented in Table \ref{tab:evaltable}.

\softname{GGIR} achieves a moderate level of maturity in \focusa{Software Project Management} but requires further improvement in \practice{code quality} and \practice{security} practices to achieve higher maturity. \softname{GGIR} lacks a structured community and is developed through voluntary efforts~\cite{ggir_website}. It shows a high level of maturity in \focusa{Software Adoptability}, indicating well-defined methodologies in this focus area.

\softname{ESMValTool} also achieves a moderate level of maturity in Software Project Management, due to its lack of adherence to security practices. However, being 
an open-source community-developed tool for climate science, it excels in \focusa{Community Engagement} and \focusa{Software Adoptability} and has implemented numerous practices in these focus areas to improve its capabilities.

Thus, using this model, users can evaluate their research software project management by identifying areas that need improvement and practices that need to be followed to achieve higher maturity. For example, in the case of \softname{GGIR},  evaluating the code quality and security capability within the \focusa{Software Project Management} focus area reveals the absence of several key practices, such as \practice{Use crash reporting}, \practice{Conduct security reviews}, \practice{Define code coverage targets}, and \practice{Follow industry standard for security}. This currently prevents \softname{GGIR} from achieving a higher maturity level. 

Van Nieuwpoort and Katz~\cite{roles2023} noted, however, the importance of roles and context in research software development, stating that not all software is created equal, and thus, not all software must follow the same practices. That is, research software should simply be mature enough to be fit for its intended purpose. 
%
In this light, it is understandable that both \softname{GGIR} and \softname{ESMvalTool} do not adhere to extensive security practices. 
These tools consists of scripts and libraries designed to support data analysis in their respective research fields, and are not typically used in an online environment, or with sensitive data. Thus, an additional refinement is necessary in future versions of \nameMM{} to take this into account.

\section{Conclusion and Future work}\label{conculsion}
In this paper we have presented 
the maturity model \nameMM{}, which provides a comprehensive framework for research software project management: \nameMM{} combines best practices from 3 different areas, namely, software engineering practices (such as requirements analysis, code quality, and security), open-source software development practices (including code review, public repository storage with version control, and community building) and best research software practices (such as making code citable and findable). To that end, we have collected components of RSMM using systematic literature review, categorized them and further refined the model through experts interviews.

 \nameMM{} is aimed at helping research and research support organizations to systematically improve their management processes and achieve desired maturity in their research software projects. \nameMM{} is a unique model that fits the needs of researchers, research software engineers, funders, policy makers, and organizations that produce research software alike as we uncovered in our interviews with research software engineers, researchers, and project managers.
Using \nameMM{}, a researcher, a research software engineer or a project manager can identify areas for improvement to achieve higher maturity levels, thereby improving their research project management. Our model allows for benchmarking project management processes across different projects and
can help making informed decisions about project funding. Funders can use this evaluation results to decide which projects to support. Additionally, policy makers can utilize the model to update and improve policies based on evaluations of previous projects.
\\
\emph{Future work:} As a final step, we will conduct case studies to evaluate the applicability of \nameMM{}. In this, we will assess the maturity of 50 research software projects, examining the relationships between concepts such as FAIR principles, quality, maturity, and impact — different but related aspects of research software. Additionally, we aim to refine our maturity model based on feedback from the experts. Lastly, we will investigate  software verification practices, which are important for software and research quality. 

\section*{Acknowledgement}
We want to acknowledge the experts who contributed to redefining \nameMM{} (not in any particular order): Bernadette Fritzsch, Jayesh Badwaik, Michael Schlottke-Lakemper, Pablo Lopez-Tarifa,  Raoul Schram, Nicolas Renaud, Arend Rensink, Jan Philipp Dietrich, Axel Loewe, Aljen Uitbeijerse, Tomas Turner-Zwinkels, and Martine de Vos. 

 
\balance
\bibliographystyle{myIEEEtran}
\bibliography{references}

\end{document}